\documentclass{elsart}

\usepackage{graphicx,amssymb}
\usepackage{epsfig,amsfonts}
\usepackage{amsmath}
\journal{pla}

\begin{document}
\begin{frontmatter}

\title{Decoherence and localization in the double well model}
\author{Gabriela Barreto Lemos}

\corauth{Corresponding author}
\ead{gabibl@fisica.ufmg.br}

\address{Dept. de F\'{\i}sica, ICEx, UFMG, CP 702, Belo Horizonte,
30123-970, Brazil}

\author{M. F. Santos}
\address{Dept. de F\'{\i}sica, ICEx, UFMG, CP 702, Belo Horizonte,
30123-970, Brazil}

\author{J. G. Peixoto de Faria}
\address{Dept. de Ci\^{e}ncias Exatas e Tecnol\'{o}gicas,
Universidade Estadual de Santa Cruz, CEP 45662-000, Rodovia Ilh\'{e}us-Itabuna
km 16, Ilh\'{e}us, Bahia, Brazil}

\author{M. O. Terra Cunha}
\address{Dept. de Matem\'atica, ICEx, UFMG, CP 702, Belo Horizonte,
30123-970, Brazil}

\author{M. C. Nemes}
\address{Dept. de F\'{\i}sica, ICEx, UFMG, CP 702, Belo Horizonte,
30123-970, Brazil}

\begin{abstract}
We use a spin-1/2 model to analyze tunnelling in a double well system coupled
to an external reservoir. We consider different noise sources such as
fluctuations on the height and central position of the barrier and propose an
experiment to observe these effects in trapped ions or atoms.
\end{abstract}

\end{frontmatter}

\maketitle

Tunnelling is a purely quantum phenomenon where a trapped massive particle
escapes from the trapping potential even though its total energy is smaller
than the trapping barrier itself. This mechanism is another manifestation of
interference in quantum mechanics. It is also the origin of
delocalization in many different quantum systems.

For example, consider a particle trapped in a symmetric double well potential,
and consider it initially localized in the left side of the barrier. According
to classical mechanics, if the energy of the particle is smaller than the
energy of the middle barrier, the particle will remain on the left side
forever. Quantum mechanics, however, predicts that if the barrier is not
infinite, then there is a non-null probability that the particle will
eventually tunnel to the right side of the trap. In fact, if the whole system
is isolated, and the particle is in a low-energy level, the dynamics of this
trapped particle will present oscillations between the localized states to the
left and right of the middle barrier and in general the particle will be
delocalized in the two sides of the barrier.

On the other hand, we also know that interference of quantum states can be
irreversibly destroyed when the corresponding system interacts and entangles
with much larger ones, so-called reservoirs, in a quantum effect known as
decoherence. This effect has been widely studied in many different systems and
it has helped the comprehension of fundamental questions such as localization
and the quantum-classical transition \cite{two}, \cite{three}, \cite{giulini},
\cite{four}, \cite{terra}, \cite{bacana}.

In this paper we study the problem of tunnelling, decoherence and localization
in the double well potential. We were motivated by the question of finding a
simple qualitative model to determine the effect of the coupling of a quantum
system, able to undergo tunnelling between two potential wells, to dissipative
degrees of freedom. We map our problem into that of a spin $1/2$ particle
coupled to a classical static magnetic field and subjected to spin-flip and
dephasing reservoirs. We show how this noise ends up localizing the particle in
one side of the barrier, and we propose an experiment to observe these effects
on trapped ions or atoms.

The manuscript is organized as follows: first, we map out the movement of the
center of mass of the trapped particle into a spin-$1/2$ problem. Then we
describe the dynamics of this fictitious spin-$1/2$ when subjected to different
noise sources to which we associate fluctuations in the position and
height/width of the separating potential barrier. Finally, we suggest an
experiment to observe this dynamics in trapped ions.

\begin{figure}
\centerline{\psfig{figure=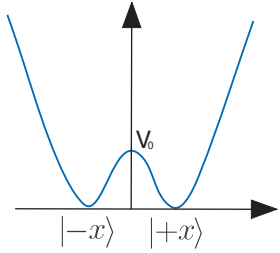}} \caption{The symmetric double well.
We associate each side of the potential barrier through which the particle can
tunnel with the eigenstates of the $S_x$ component of the spin.}
\end{figure}%

Let us consider a particle in a symmetric double well potential, i.e. an
infinite one-dimensional well ($V(-x_{max})=V(x_{max})=\infty$) with a finite
potential barrier in the center ($0 < V(0) = V_0$). Due to the symmetry of this
configuration, the eigenstates of this confining Hamiltonian can be either even
(symmetrical) or odd (antisymmetrical) functions. Consider the central
potential barrier very high; the arrangement can be looked upon as two infinite
potential wells. In such a case each possible energy value is twofold
degenerate. When the finite height of the potential barrier is taken into
account, the tunnel effect across the barrier removes the degeneracy, and the
first levels splits, giving rise to doublets $\{|n_s\rangle, |n_a\rangle\}$
($s$ for symmetrical, $a$ for antisymmetrical), with respective energies
$E_{n_s}$ and $E_{n_a}$. The large barrier assumption can be written $E_{n_s},
E_{n_a} << V_0$ for small $n$. It also implies $|E_{n_s}-E_{n_a}|\ll
|E_{n+1_s}-E_{n_s}|$, i.e. the energy separation between neighbor doublets is
much larger than their splitting. This allows us to treat each doublet as a
separate two-level system.

From now on, we will consider only the first doublet and represent it as a
spin-$1/2$ particle precessing around an uniform classical magnetic field in a
fictitious spin picture~\cite{1}. There are infinite equivalent ways of
building this analogy. We will choose to associate each side of the potential
barrier with the eigenstates $|-x \rangle$ (left side of the barrier) and $|+x
\rangle$ (right side of the barrier) of the $S_x$ component of this spin. In
this case, the symmetrical and antisymmetrical stationary states in the well,
given by $|s \rangle = \frac{|+x \rangle + |-x \rangle}{\sqrt{2}}$ and $|a
\rangle = \frac{|+x \rangle - |-x \rangle}{\sqrt{2}}$, will be associated with
the eigenstates of $S_z$ ($\sigma_z = |s\rangle\langle s| - |a\rangle \langle
a|$). The back and forth motion of the particle through the potential barrier
inside the well can be associated with the precession of this fictitious
spin-1/2 around an uniform magnetic field parallel to the $O_z$ axis, with
frequency $\omega= \frac{|E_{0_s}-E_{0_a}|}{\hbar}$. This oscillation can also
be viewed as a succession of constructive and destructive interference of the
symmetrical and antisymmetrical wavefunctions. The dynamics of this isolated
system is then described by the Hamiltonian $H = -\frac{\omega}{2} \sigma_z$.

Considered as an open quantum system, the dynamics of this virtual two-level
particle is better described by the following Lindblad master equation
\cite{2}(in units of $\hbar = 1$):
\begin{equation}
\dot{\rho} = -i \frac{\omega}{2}
[\sigma_z,\rho]-\frac{1}{2}\sum_{k=1}^n \{\Gamma_k^\dag \Gamma_k
\rho + \rho \Gamma_k^\dag \Gamma_k - 2 \Gamma_k \rho \Gamma_k^\dag
\},
\end{equation}
with each $\Gamma_k$ being associated to a different decohering
process.

Continuing our analogy, we can associate random fluctuations of
the height or width (with symmetry preservation) of the barrier to
a phase reservoir in the spin system represented by
$\Gamma_1=\sqrt{k_1} \sigma_z$. We justify this equivalency by the
following argument: small fluctuations in the height or width of
the barrier (i.e. vertical or horizontal symmetric shaking of the
barrier) do not change the eigenstates of the barrier, but only
result in small fluctuations in $\omega$, causing dephasing
between the symmetrical and antisymmetrical states, which
characterizes a phase reservoir. In the magnetic field analogy,
this corresponds to fluctuations on $B_z$, but keeping
$B_x=B_y=0$.

Another decoherence source can be associated with fluctuations of
the position of the barrier. Small shifts around the central
position of the barrier result in new eigenstates for the system.
Those can be obtained as linear combinations of the former ones,
and the whole process can be described as a small $B_x$
fluctuation, for example. We can therefore associate this
phenomenon with a spin flip reservoir described by
$\Gamma_2=\sqrt{k_2} \sigma_x$.

We first consider only the phase reservoir $\Gamma_1$, i.e. the shaking of the
potential barrier in the well. The master equation describing the evolution of
the system becomes:
\begin{equation}
\dot{\rho} = -i \frac{\omega}{2}
[\sigma_z,\rho]-k_1\{\rho-\sigma_z \rho \sigma_z \}, \label{difme}
\end{equation}
since $\sigma_z = \sigma_z^\dag$  and $\sigma_z^2=\mathbb{I}$. To solve the
above master equation we consider the Bloch vector parametrization
\begin{equation}
\rho(t) = \frac{1}{2} \begin{pmatrix}
  1+S_z(t) & S_x(t)+i S_y(t) \\
  S_x(t)-i S_y(t) & 1-S_z(t)
\end{pmatrix},
\end{equation}
with general pure state initial condition (spin in an arbitrary direction)
\begin{equation}
\rho(0) = \frac{1}{2} \begin{pmatrix}
  1+\cos{\theta} & e^{-i\phi} \sin{\theta} \\
  e^{i\phi} \sin{\theta} & 1-\cos{\theta}
\end{pmatrix}.
\end{equation}
The equations of motion for $S_x$, $S_y$, and $S_z$ obtained are:
\begin{eqnarray}
\dot{S}_x & = & -  2k_1S_x + \omega S_y, \nonumber \\
\dot{S}_y & = & -  \omega S_x - 2k_1S_y, \nonumber \\
\dot{S}_z & = & 0.
\end{eqnarray}
As one should expect, the phase reservoir preserves the $S_z$ component of the
spin, which means that the eigenstates of $S_z$ are stationary solutions of
equation~\eqref{difme}. In general, the solution to the master
equation~\eqref{difme} is given by
\begin{eqnarray}
S_x(t) & = & e^{-2k_1 t} (A\cos{\omega t} + B \sin{\omega t}), \nonumber\\
S_y(t) & = & e^{-2k_1 t}(-A\sin{(\omega t)} + B \cos{(\omega t)}), \nonumber \\
S_z(t) & = & C,
\end{eqnarray}
with the constants $A$, $B$ and $C$ determined by the initial state. The time
dependent density matrix is then given by
\begin{equation}
\rho(t) = \frac{1}{2} \begin{pmatrix}
  1+\cos \theta & e^{-i\phi}\sin \theta e^{-2(k_1 + i\omega)t} \\
  e^{i\phi}\sin \theta e^{-2(k_1 - i\omega)t} & 1-\cos \theta
\end{pmatrix}.
\label{mt1}
\end{equation}
For example, let us consider an initial state $|+x\rangle$ ($\theta = \pi/2$
and $\phi=0$), which corresponds to an even superposition of the symmetrical
and antisymmetrical eigenstates of the barrier. In this case, the particle is
initially localized in the left side of the barrier. Note, however, that this
localization should be viewed as constructive interference of the symmetric and
antisymmetric solutions of the trapping potential. Also note that, once the
system evolves in time, the particle will tunnel through the barrier and the
probability to find it on the left will be given by
\begin{equation}
P_l(t)=\frac{1}{2}\left[1+e^{-2k_1t}\cos \left(\omega t\right)\right].
\end{equation}
The above probability shows that the particle undergoes quantum dissipative
oscillations between both sides of the barrier, but, asymptotically tends to
localize, now in the classical statistical sense, with equal probabilities
$P_l(\infty)=P_r(\infty)=1/2$, either to the left or the right of the central
barrier. By classical localization we mean that the initial pure state turns
into a statistical mixture, i.e. it is no longer possible to observe
interference in the system

We can also observe this classical localization by calculating the purity of
the system as a function of time, which, for two level systems, can be defined
by
\begin{equation}
\zeta \left(t\right) = 2 \textrm{tr} \rho (t)^2 - 1.
\end{equation}
This quantity varies from zero, for the completely random state, to one, for
pure states. In fact, it gives the squared norm of the Bloch vector. From Eq.
(7), one obtains
\begin{equation}
\zeta \left( t\right) = e^{-4k_1t},
\end{equation}
which confirms that the quantum state of the particle evolves from a pure state
($\zeta (0) =1$) to a classically localized state ($\zeta (t \gg 1/k_1)
\rightarrow 0$) due to the action of the external reservoir.

Also note that for an arbitrary angle $\theta$, the density matrix for $t \gg
k_1$ tends again to a mixture of states $|+_z\rangle$ and $|-_z\rangle$, but
now with different populations.
\begin{equation}
\rho(t\rightarrow \infty) = \frac{1}{2} \begin{pmatrix}
  1+\cos \theta & 0 \\
  0 & 1-\cos \theta
\end{pmatrix}.
\label{mt1}
\end{equation}
This result should be expected given that the phase reservoir
affects only the coherence of the quantum state, i.e. it preserves
the original populations of the quantum state in the $\sigma_z$
basis.

\begin{figure}
\centerline{\psfig{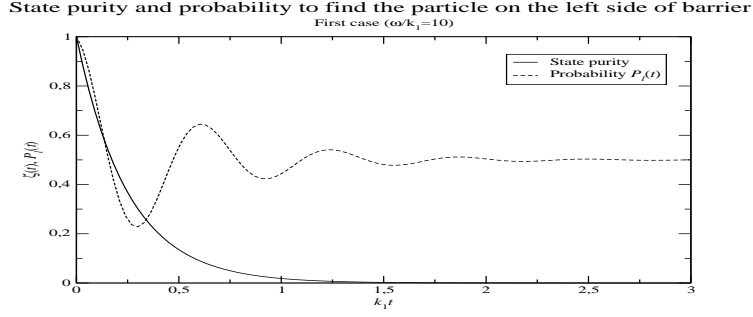}}
\caption{The tunnelling system coupled to the $\sigma_z$ reservoir: state
purity $\zeta (t)$ and probability to find the particle in the left side of the
double well $P_l(t)$ as a function of time, when the particle is initially
prepared in the same left side ($\theta=\frac{\pi}{2}$ and $\phi=0$) and for
$k_1=1$ and $\omega=10$. Note that the purity of the system tends to zero,
indicating loss of coherence, and the probability tends to one-half, thus
indicating localization.}
\end{figure}

\begin{figure}[h]
\begin{center}
\includegraphics[width=2in]{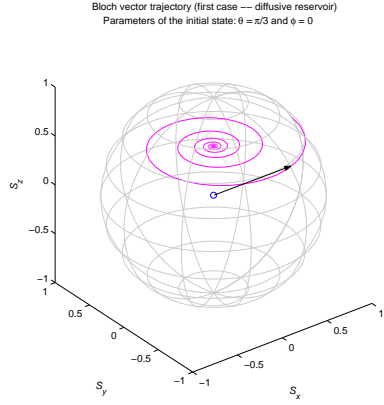}
\end{center}
\caption{A schematic diagram of the Bloch vector evolving in the Bloch sphere
for an arbitrary initial state defined by the angles $\theta$ and $\phi$. Only
the $x$ and $y$ components of the Bloch vector are affected by the reservoir.
On the other hand, the $z$ component is preserved and, in the asymptotic limit,
the Bloch vector tends to the projection of the initial state on the $O_z$
axis.} \label{BLOCH2}
\end{figure}

A Bloch sphere picture allows a simple visual comprehension of these effects
(see figure ~\ref{BLOCH2}). First, one should remember that a Bloch vector only
describes density operators. It can be thought of as an ensemble average on
pure state realizations. The isolated and initially pure system would imply
circular trajectories with constant $z$ component and constant angular
velocity. The phase reservoir induces velocity fluctuations. The ensemble then
spreads over the circle, and the $x$ and $y$ components of the Bloch vector
decay exponentially. For long times (compared to $k_1^{-1}$) the vector will be
very close to the projection of the initial state onto the $O_z$ axis.

It is interesting to notice that two special states are not affected by this
reservoir. The $|+_z\rangle$ and $|-_z\rangle$ states can be viewed as rigorous
pointer states~\cite{3} since they never lose purity.

\begin{figure}[t]
\begin{center}
\includegraphics[height=5cm]{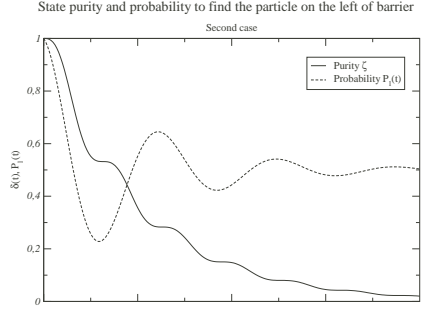}
\end{center}
\caption{The tunnelling system coupled to the $\sigma_x$ reservoir: purity
$\zeta(t)$ and probability $P_l(t)$ as a function of time, when the particle is
initially prepared in the left side of the potential well, for $k_2=1$ and
$\omega =10$. One clearly notes two distinct dynamics in the purity. The
plateaux occur at moments in which the system is at one side of the potential
well (near integer and half-integer tunnelling periods).} \label{graph3}
\end{figure}

Next we consider only the spin-flip reservoir, which corresponds to
fluctuations of the position of the barrier inside the potential well (i.e. an
asymmetric disturbance). The corresponding master equation is given by:
\begin{equation}
\dot{\rho} = -i \frac{\omega}{2}
[\sigma_z,\rho]-k_2\{\rho-\sigma_x \rho \sigma_x \},
\label{flipme}
\end{equation}
since $\sigma_x = \sigma_x^\dag$ and $\sigma_x^2 = \mathbb{I}$. In
this case, the equations of motion for $S_x$, $S_y$ and $S_z$ are:
\begin{eqnarray}
\dot{S}_x & = & \omega S_y, \nonumber \\
\dot{S}_y & = & -2k_2 S_y -\omega S_x, \nonumber \\
\dot{S}_z & = & -2 k_2 S_z.
\end{eqnarray}
The solution to equation \eqref{flipme} is:
\begin{equation}
\rho(t) = \frac{1}{2} \begin{pmatrix}
  1+e^{-2k_2 t}\cos \theta  & c^*(t) \\
  c(t) & 1-e^{-2 k_2 t}\cos \theta
\end{pmatrix},
\end{equation}
with
\begin{eqnarray*}
c(t) = e^{(-k_2 t)}\sin(\theta)\left\{\exp(-i\phi) \cos \epsilon t + \frac{\sin
\epsilon t}{\epsilon} [k_2 \exp(i\phi) - i\omega \exp(-i\phi)]\right\},
\nonumber \cr
\end{eqnarray*}
where $\epsilon^2 = \omega^2 - k_2^2$. Again, the interaction with the
reservoir destroys the coherence in the initial quantum state, which is
transformed into a complete statistical mixture for $k_2 t\gg 1$. As before,
the asymptotic state is also localized in the classical statistical sense.
However, since the spin flip reservoir acts exactly by random population
transfer, the final state presents equal populations on both sides of the
barrier, independently of the initial state.

If the system is initially prepared in state $|+x\rangle$ ($\theta = \pi/2$ and
$\phi=0$)), once it evolves in time, the particle will tunnel through the
barrier and the probability to find it on the left side, will be given by
\begin{equation}
P_l(t) = \frac{1}{2}\left(1 + e^{-k_2t}\right)(\cos\epsilon t
+\frac{k_2}{\epsilon} \sin\epsilon t),
\end{equation}
and the purity $\zeta$ of the system evolves in time according to
\begin{equation}
\zeta (t)= \left| c\left( t\right)\right| ^2 + e^{-4k_2t} cos^2 \theta.
\end{equation}
In figure~\ref{graph3} one clearly notes two different dynamics in
the time evolution of the purity of the quantum state of the
particle. When the state of the particle is more localized at one
of the sides, the particle loses purity in a much slower rate,
while when the particle is tunnelling through the barrier its
quantum state loses purity in a faster rate. This behavior is
indeed expected given that the particle's position wave function
feels the fluctuations of the barrier much strongly when crossing
it.

In this case, the Bloch vector picture is subtler. Again, the isolated system
would rotate aroun the $O_z$ axis, but now the fluctuations vary the axis of
the movement. The ensemble spreads to neighbor circles, and the circle plains
also fluctuate. In the asymptotic limit ($k_2t \gg 1$) all points are possible
and the Bloch vector tends to the center of the sphere.

\begin{figure}
\begin{center}
\includegraphics[width=2in]{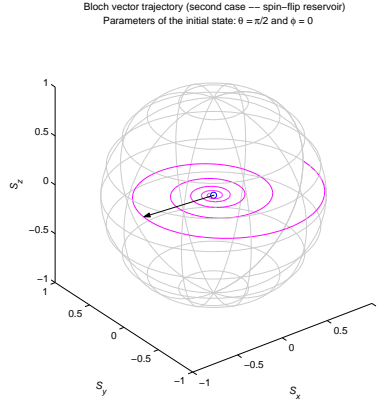}
\end{center}
\caption{When the tunnelling system is coupled to a spin-flip reservoir, no
Bloch vector remains fixed. Here we show one of the pointer states, aligned
with the $O_x$ axis. Even this state is not preserved because it is taken off
its original axis by the unitary evolution, making it vulnerable to the
reservoir and allowing decoherence to take place.} \label{BLOCH3}
\end{figure}

With the usual assumption $\omega \gg k_2$, the $|\pm_z\rangle$ states are the
most robust states. They can also be classified as pointer states, by the
predictability sieve criterion~\cite{3}, but in a sense different to that of
the first case. Here, no vector is rigorously unaffected. The fluctuations
induce random rotations around $O_x$ axis so that when the particle is
localized on one side of the barrier, its Bloch vector being aligned with
$O_x$, it is \emph{instantaneously} immune to the reservoir. This means that
the spin-flip reservoir would tend to keep $|\pm_x\rangle$ states fixed,
however the unitary evolution takes them off the $O_x$ axis, and the other
vectors that appear in their trajectories are strongly affected by decoherence
(see fig. ~\ref{BLOCH3}).

Finally, we will describe a feasible trapped ions experiment to observe the
decoherence effects on double well tunnelling. The basic ingredients are: a
single trapped ion, two of its electronic levels, $\left \{|g\rangle,
|e\rangle\right \}$, and an external laser field of frequency $\omega_L$. The
ion is initially prepared in lower level $|g\rangle$ and it is cooled down to
its vibrational ground state which, for a harmonic trap, is a gaussian
distribution around the center of the trap. The external laser field is then
focused to the right of the ionic distribution, and it is largely blue shifted
in respect to the ionic $|g\rangle \rightarrow |e\rangle$ transition, i.e.
$\Delta = \omega_{L}-\omega_{eg} \gg G$, where $G$ is the coupling constant of
the chosen transition. The dispersive interaction between the ion and the laser
beam induces a repulsive dipole optical force in the ion, pushing it away from
the focus of the laser field. In fact, this new interaction constitutes an
optical potential barrier to the motion of the ion. The laser is then pushed to
the left until it reaches the center of the trapping potential, configuring a
double well potential for the motion of the ion. If the process is adiabatic,
the ion is pushed to the left side of the well and if the laser is intense
enough, the ion remains there, i.e. there is no tunnelling to the right side of
the trap. If the laser intensity is now reduced, the central barrier is lowered
and the ion tunnels through this barrier undergoing Rabi oscillations between
the left and right sides of the double well. Fluorescence of the ion can then
be used to measure its position in either side of the well for different times.
The fluctuations of the dimensions and positions of the barrier can be easily
obtained in a controlled way by randomly altering the intensity or the position
of the laser beam.

In this letter we used a simple model to show that, for long times, a quantum
system, able to undergo tunnelling between two potential wells, coupled to
dissipative reservoirs tends to localize. We also suggested a simple experiment
with trapped ions to observe this effect.

\begin{figure}[h]
\begin{center}
\includegraphics[width=2in]{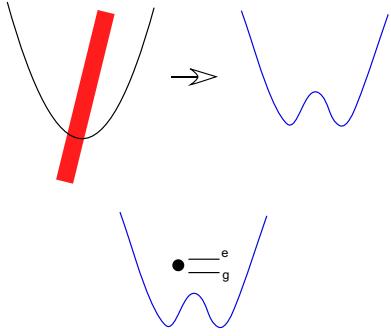}
\end{center}
\caption{A schematic draft of an ion trap traversed by an external
laser field forming a double potential well.} \label{diagrama}
\end{figure}

MCN and MFS thank CNPq for financial support. GBL thanks CNPq and
CAPES for financial support. JGPF thanks to PROPP-UESC for
financial support.





\begin{thebibliography}{9}
\bibitem{two}
A. O. Caldeira and A. J. Leggett, Ann. Phys. (NY) \textbf{149}, 374 (1983).
\bibitem{three}
A. J. Leggett \textit{et al.}, Rev. Mod. Phys. \textbf{59}, 1 (1987).
\bibitem{giulini}
D. Giulini et al., \emph{Decoherence and the Appearance of a Classical World in
Quantum Theory}, Springer-Verlag, Telos (1996).
\bibitem{four}
A. N. Salgueiro, A. F. R. de Toledo Piza, J. G. Peixoto de Faria
and M. C. Nemes, Phys. Rev. A \textbf{64}, 032113 (2001).
\bibitem{terra}
M. O. Terra Cunha and M. C. Nemes, Phys. Lett. A \textbf{329}, 409 (2004).
\bibitem{bacana}
W. H. Zurek, Physics Today \textbf{44}, 36, (1991).
\bibitem{1}
C. Cohen-Tannoudji, B. Diu et F. Lalo\"e, {\it M\'ecanique Quantique}, Vol. 1,
Hermann, France (1977).
\bibitem{2}
G. Lindblad, Com. Mat. Phys. \textbf{48}, 119, 1976.
\bibitem{3}
W. H. Zurek, Prog. Theor. Phys., \textbf{89}, 281, 1993.

\end{thebibliography}
\end{document}